\documentclass[aps,pra,twocolumn,superscriptaddress,showpacs,floatfix]{revtex4-2}
\usepackage{graphicx}
\usepackage{dcolumn}
\usepackage{bm}
\usepackage{hyperref}
\usepackage{comment}
\usepackage{xcolor}   
\usepackage{amsmath}
\usepackage{float}
\usepackage{amsfonts}

\usepackage[normalem]{ulem}
\usepackage{xcolor}
\usepackage{soul}

\begin{document}

\preprint{APS/123-QED}

\title{Heralded Induced-Coherence Interferometry in a Noisy Environment}%

\newcommand{\IISc}{Quantum Optics and Quantum Information, 
Department of Electronic Systems Engineering, 
Indian Institute of Science, Bengaluru 560012, India.}

\author{L. Theerthagiri }
\email{tgiri9157@gmail.com}
\affiliation{\IISc}
\author{Balakrishnan Viswanathan}
\affiliation{Optics and Quantum Information Group, The Institute of Mathematical Sciences, 
C. I. T. Campus, Taramani, Chennai 600113, India.}

\author{C. M. Chandrashekar}
\affiliation{\IISc}

\date{\today}

\begin{abstract}
Induced-coherence interferometry, first introduced in the Zou–Wang–Mandel (ZWM) setup, enables retrieval of object information from the interference pattern of light that never interacted with the object. This scheme relies on two identically correlated photon pairs and the absence of “which-way” information about the photons illuminating the object to induce coherence in their companions. In pervious studies, effect of thermal background on ZWM interferometer was considered and here we explicitly include background noise and analyze the interference visibility in both low- and high-gain regimes, revealing how thermal photons introduce an incoherent offset that lowers the observed interference contrast. We show that the visibility can be restored either by optimal attenuation or by extending the geometry to a three-SPDC configuration. Furthermore, we demonstrate 
that introducing heralded detection removes the detrimental effect of thermal background noise, restoring high-contrast interference fringes.
\end{abstract}

\maketitle

\section{Introduction}
Interferometry based on path indistinguishability—first demonstrated by Zou, Wang, and Mandel (ZWM)~\cite{Zou1991}—shows that aligning the idler modes produced through spontaneous parametric down-conversion (SPDC) from two identical sources can erase the which-way information of the idler photons, thereby inducing first-order coherence between the corresponding signal fields, even when the idlers are not detected, as shown in Fig.~\ref{2psdc}. This erasure of path identity forms the basis of a broad class of \emph{induced-coherence} techniques, including imaging with undetected photons (IUP)~\cite{Lemos2014} and studies of complementarity and the spontaneous–stimulated crossover~\cite{HeuerPRL2015,MilonniPRA2020}. 

 The ZWM interferometer operates on the principle of ``path indistinguishability'' or the absence of ``which-way'' information \cite{zou,zou1991induced}. The erasure of the path identity of the idler photons through their alignment induces coherence in the companion, signal photons. The injection of thermal photons in the system can disturb the alignment of the idler modes which in turn affects the induced coherence in signal photons. The presence of background noise in the interferometer manifests through the reduced contrast in the interference fringes of the signal field. The intensity of the signal measured by the detector then acquires an incoherent pedestal proportional to the strength of the noise, which grows with background brightness and obscures the phase-sensitive term.

At thermally bright bands (mid-IR, THz, or microwave) or under deliberate noise injection, the induced-coherence picture breaks down. In the ZWM geometry, an object in the idler arm acts as a lossy beam splitter with transmissivity \(T\) that mixes the idler with a thermal mode of mean photon number \(N_B\). The singles signal \(I_s(\phi)\) then acquires an incoherent pedestal proportional to \((1-T)N_B\), which grows with background brightness and obscures the phase-sensitive term. Similar thermal-background effects were analyzed by Ma \emph{et al.}~\cite{Phillips}, who showed that imaging with undetected photons is largely immune to weak thermal seeding. In the low-gain regime, however, the singles visibility rapidly collapses as thermal photons dominate the background, while in the high-gain regime a finite contrast persists but decreases monotonically with increasing \(N_B\).

Singles visibility can be partially recovered by passive balancing---either by \emph{optimal attenuation} of the stronger signal arm or by employing a \emph{three-SPDC} configuration that restores balance intrinsically at higher gain---both achieving the coherence-bound visibility \(|g^{(1)}_{12}|\) without coincidences. In the strongly thermal, low-gain regime where these passive methods fail, we next introduce \emph{heralded induced-coherence interferometry} as an active quantum-filtering approach \cite{sperling,Somaraju,MandelWolf1995}. Heralded detection projects the signal onto the correlated two-photon subspace, removing the thermal pedestal and yielding visibility and signal-to-noise ratio independent of \(N_B\). This conditional-interference regime, unexplored in previous ZWM or quantum-induced-coherence (QuIC) LiDAR implementations, establishes heralding as a robust route to noise-resilient quantum interferometry.

The key physical advantage of this heralded ZWM configuration is that, the conditional (idler-heralded) measurement projects out all uncorrelated thermal photons, thereby restoring high-contrast interference even in thermally bright environments where the visibility from the singles degrades. Conditioning projects measurements onto the two-photon (pair) sector, so uncorrelated thermal photons — while they elevate singles' rates — do not contribute to the phase-carrying pair correlations that set the fringe amplitude. More specifically, the conditional measurement enhances the visibility in the low-brightness regime. In the ideal (orthogonal) single-mode, zero-accidental limit, the resulting \emph{heralded visibility} becomes insensitive to $N_B$. 
In practice, imperfections such as finite detection efficiency, dark counts, and mode mismatch would reduce the observed visibility, yet the fundamental noise-rejection mechanism provided by heralding remains unchanged.

Our approach is distinct from two related paradigms: \emph{quantum illumination (QI)} and \emph{quantum-induced-coherence LiDAR (QuIC-LiDAR)}. Unlike QI, which leverages signal–idler correlations for binary target detection in bright, lossy environments 
(\text{low reflectivity} $\kappa \!\ll\! 1$, \text{weak transmission} $N_S \!\ll\! 1$, \text{bright background} $N_B \!\gg\! 1$) 
and is quantified by error exponents or ROC \cite{giri} curves rather than first-order fringe visibility 
\cite{Lloyd2008,Tan2008,Pirandola2018,Shapiro2020}, 
our analysis focuses on induced-coherence interferometry and does not assume or claim QI-style error-exponent advantages. 
Similarly, QuIC-LiDAR exploits induced coherence for ranging without directly detecting the probe beam 
\cite{Qian2023}; our analysis provides a complementary route to robustness—via heralding—when induced-coherence sensing is pushed into thermally bright regimes.

The remainder of this paper is organized as follows. Section~II derives singles behavior and visibility under thermal injection. Section~III analyzes attenuation and its impact on contrast. Section~IV quantifies the \emph{signal-to-noise ratio (SNR)} limits for unconditional detection. Section~V develops the heralded-measurement theory with realistic detectors and presents the resulting visibilities and SNR scalings. We conclude in Section~VI with implications for noise-resilient induced-coherence sensing.

\section{Induced coherence interferometer: Theoretical framework}

    \begin{figure}	
		\includegraphics[width=0.49\textwidth]{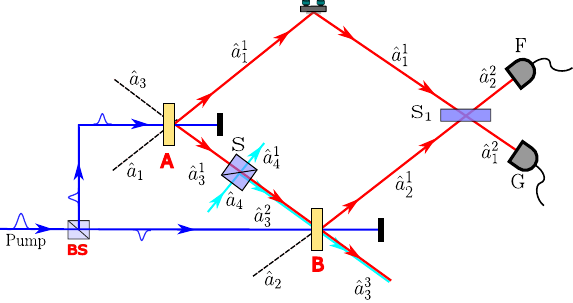}
		\caption{Experimental setup for observing induced coherence. Two coherently pumped nonlinear crystals (A, B) generate 
signal  ($\hat{a}_{1}^{1},\hat{a}_{2}^{1}$)-idler pairs  ($\hat{a}_{3}^{1},\hat{a}_{3}^{3}$). The idler modes are aligned into a common path 
that includes a filter $S$ with a variable transmittance $T$ (modeled as a beam splitter) that both attenuates the idler and injects a 
thermal mode with a mean photon number $N_B$. The signal modes are 
combined at a beam splitter $S_1$ and detected at outputs F and G. 
The induced coherence in crystal B is thus controlled by attenuation 
and thermal injection in the shared idler channel.}
\label{2psdc}
	\end{figure}

The notion of induced coherence was first introduced to demonstrate single-photon interference by path indistinguishability using the ZWM interferometer \cite{zou1991induced,Zou1991}. Over the past decade, this setup has been used to implement practical tasks such as quantum imaging with undetected photons \cite{Lemos2014, lahiri2015theory} where the light that illuminates the object is not detected. The image is constructed from the interference pattern of the companion photon that never interacted with the object. There is no coincidence detection in this technique. As a result, this scheme is well suited to image and detect objects at wavelengths for which efficient time-resolving detectors may not be available yet. Although this technique of imaging was first studied using entangled photon pairs, subsequently, it has been explored in the classical regime as well \cite{shapiro2015classical,cardoso2018classical}.

In many of the investigations involving the ZWM interferometer, so far, thermal noise in the background has been ignored. Here, we explicitly include thermal noise and look at its effect on the visibility of the interference pattern in both, low- and high-gain regimes, respectively. In the low-gain regime \cite{Kolobov2017,Machado2024}, at most one pair is produced across the two coherently pumped nonlinear crystals (A,B). If A generates a pair, the signal $(\hat{a}_{1}^{1})$ is pair-correlated with the idler $(\hat{a}_{3}^{1})$ that traverses the object port (mixed with a thermal mode); If crystal B generates a pair, its signal $(\hat{a}_{2}^{1})$ and idler $(\hat{a}_{3}^{3})$ photons remain correlated through the SPDC process, and the idler $(\hat{a}_{3}^{3})$ photon itself does not interact directly with the thermal mode. However, because the idler $(\hat{a}_{3}^{2})$ input to crystal B contains a small thermal admixture (vacuum–thermal induced coherence), both the signal $(\hat{a}_{2}^{1})$ and idler $(\hat{a}_{3}^{3})$ fields inherit a weak, classical (intensity-level) correlation with the thermal background. This residual coupling gives rise to an additive background term in the singles intensity, while no phase-sensitive or quantum correlation exists with the thermal photons. The idlers ($\hat{a}_{3}^{2},\hat{a}_{3}^{3}$) from A and B are aligned into a common spatial mode, rendering the signal ($\hat{a}_{1}^{1},\hat{a}_{2}^{1}$) origin indistinguishable and producing interference at the final beam splitter.

We model the object as a lossy beam splitter $S$ with intensity transmittance $T\!\in[0,1]$ that mixes the idler with a thermal mode of mean photon number $N_B$ \cite{Phillips,qian}, and combine the signal photons at a $50{:}50$ beam splitter $S_1$ (Fig.~\ref{2psdc}).The singles intensities at the two signal outputs, $N_\pm(\phi)$, oscillate with phase $\phi$ and define a fringe visibility $\mathcal V=(N_{\max}- N_{\min})/(N_{\max}+N_{\min})$. Working at low gain with Bogoliubov parameters $V_j=|v_j|^2$ and $|u_j|^2-|v_j|^2=1$, the singles at the two outputs (detectors $F$ and $G$ in Fig.~\ref{2psdc}) are
\begin{equation}
\label{eq:Npm_thermal_short}
\begin{split}
N_{\pm}(\phi)=&\tfrac{1}{2}\Big[ V_A + V_B + T V_A V_B + (1-T)N_B V_B\\ 
&\pm 2\sqrt{(1+V_A)V_A V_B T}\,\cos(2\phi)\Big].
\end{split}
\end{equation}
The corresponding visibility is
\begin{equation}
\label{eq:V_general_short}
\mathcal{V}=\frac{2\sqrt{(1+V_A)V_A V_B T}}
{V_A + V_B + T V_A V_B + (1-T)N_B V_B}\,,
\end{equation}
which shows that the numerator is set by the coherent pair amplitude, while the denominator contains an additive thermal pedestal $\propto(1-T)N_B$ that degrades singles visibility in noisy regimes. For $N_B=0$, the pedestal vanishes and Eq.~\eqref{eq:V_general_short} reduces to the standard ZWM visibility \cite{Kolobov2017},
\begin{equation}
 \displaystyle \mathcal{V}=\frac{2\sqrt{(1+V_A)V_A V_B T}}{V_A+V_B+T V_A V_B}.  
\end{equation}
 \begin{figure}[t]
  \centering
  \includegraphics[width=1\linewidth]{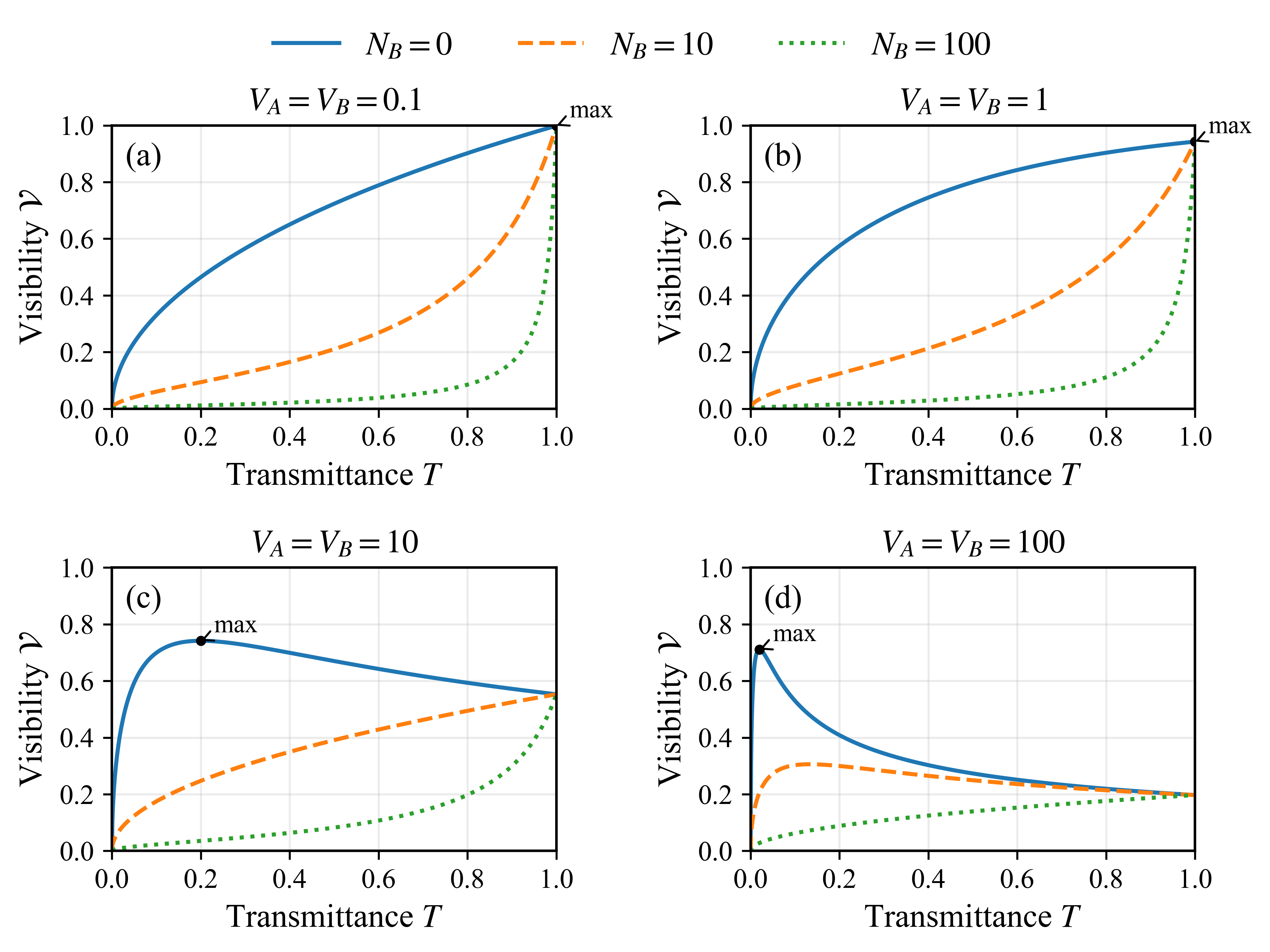}\hfill
  \caption{Singles visibility $\mathcal{V}$ versus idler transmittance $T$ for three thermal backgrounds $N_B\in\{0,10,100\}$ (blue, orange, green), computed from Eq.~\eqref{eq:V_general_short}. Each panel corresponds to the crystal gains: Low gain: (a)  $V_A=V_B=0.1$, High gain: (b) $V_A=V_B=1$, (c) $V_A=V_B=10$,
  (d) $V_A=V_B=100$. For low gain the visibility increases monotonically with $T$ and approaches the vacuum-limit curve as $N_B\!\to\!0$. At high gain the $N_B=0$ curve exhibits a maximum at finite $T$ due to competition between the $\sqrt{T}$ scaling of the numerator and the $T V_A V_B$ term in the denominator of Eq.~\eqref{eq:V_general_short}; thermal backgrounds suppress
  visibility except near $T\!\approx\!1$, where the pedestal $(1-T)N_B V_B$ vanishes.}
  \label{fig:vis_vs_T_gains}
\end{figure}

Figure~\ref{fig:vis_vs_T_gains} illustrates the dependence of the singles visibility $\mathcal{V}$ on the idler transmittance $T$ for several thermal backgrounds $N_B$. The plots, obtained from Eq.~\eqref{eq:V_general_short}, show that for weak parametric gain ($V_A,V_B\ll1$) the visibility increases monotonically with $T$ and approaches the vacuum-limit behavior as $N_B\!\to\!0$. In this regime, thermal photons merely add an incoherent pedestal $(1-T)N_B V_B$ that suppresses contrast but does not modify the functional $\sqrt{T}$ scaling of the numerator. At higher gain ($V_A,V_B\!\gtrsim\!1$), competition between the coherent $\sqrt{T}$ dependence in the numerator and the nonlinear $T V_A V_B$ term in the denominator produces a peak in visibility at finite $T$. As $N_B$ increases, this peak diminishes and the visibility curve flattens except near $T\!\approx\!1$, where the thermal pedestal vanishes.

\section{First-order coherence bound and visibility recovery}

As noted in Ref.~\cite{Kolobov2017}, the ultimate singles-visibility is limited by the first-order degree of coherence between the two \emph{signal} modes just before the
final beam splitter,
\begin{equation}
\label{eq:g12_def}
|g^{(1)}_{12}| \;=\; \frac{|\langle \hat a_1^\dagger \hat a_2\rangle|}
{\sqrt{\langle \hat N_1\rangle\,\langle \hat N_2\rangle}}\,,\qquad
\hat N_j=\hat a_j^\dagger \hat a_j.
\end{equation}
Using the Heisenberg relations in App.~\ref{apa} for the ZWM geometry with idler transmittance $T$ and a thermal mode with mean photon number $N_B$ injected at $S_1$, we obtain the pre-beam-splitter moments
\begin{align}
\label{eq:preBS_moments}
\langle \hat N_1\rangle &= V_A, \nonumber\\
\langle \hat N_2\rangle &= V_B\big[\,1 + T V_A + (1-T)N_B\,\big], \nonumber\\
|\langle \hat a_1^\dagger \hat a_2\rangle| &= \sqrt{T(1+V_A)\,V_A\,V_B}\,.
\end{align}
Hence the induced coherence (including thermal injection) is
\begin{equation}
\label{eq:g12_thermal}
|g^{(1)}_{12}|\;=\;
\sqrt{\frac{T(1+V_A)}{1 + T V_A + (1-T)N_B}}\;,
\end{equation}
which reduces to the standard ZWM result for $N_B=0$ and is independent of $V_B$.

The singles intensities at the outputs of $S_2$ obey
$N_\pm=(\langle \hat N_1\rangle+\langle \hat N_2\rangle)/2
\pm \mathrm{Re}\,\langle \hat a_1^\dagger \hat a_2\rangle$,
so the singles fringe visibility is
\begin{equation}
\begin{split}
\label{eq:V_from_moments}
\mathcal V \;=\;& \frac{2\,|\langle \hat a_1^\dagger \hat a_2\rangle|}
{\langle \hat N_1\rangle+\langle \hat N_2\rangle}\\
\;=\;&
\frac{2\sqrt{T(1+V_A)\,V_A\,V_B}}
{V_A + V_B + T V_A V_B + (1-T)N_B V_B},
\end{split}
\end{equation}
which matches Eq.~\eqref{eq:V_general_short}. Equations~\eqref{eq:g12_thermal}–\eqref{eq:V_from_moments}
clarify the trends in Fig.~\ref{fig:vis_vs_T_gains}: for large $T$ and high gain, the idler-seeded
emission in crystal B makes $\langle \hat N_2\rangle \gg \langle \hat N_1\rangle$, lowering
$\mathcal V$ well below the coherence bound.

\paragraph*{Recovering visibility at high $T$.}
Singles visibility is maximized (for fixed coherence) when the two signal arms have equal intensity at $S_1$, i.e. when $\langle \hat N_1\rangle=\langle \hat N_2\rangle$, in which case $\mathcal V_{\max}=|g^{(1)}_{12}|$. Two practical routes achieve this balance:

\emph{(i) Attenuate the stronger signal arm (from crystal B) before $S_1$ .} Optimizing over $V_B$ \cite{Kolobov2017} (or an equivalent attenuation factor in the B arm) yields
\begin{equation}
\label{eq:V2_atten_opt}
\mathcal V_{\text{2SPDC+att,opt}}(T,N_B)
\;=\;
\sqrt{\frac{T(1+V_A)}{1 + T V_A + (1-T)N_B}},
\end{equation}
which is exactly the optimized singles visibility and it equals the coherence bound.

\emph{(ii) Three-crystal (three-SPDC) variant}, 
We adopt the three-SPDC (“3SPDC”) configuration \cite{Heuer2015PRL,Heuer2015PRA}, adding a third source in the $A$-signal arm while retaining thermal injection at $S$. The resulting singles visibility will be,

\begin{equation}
\label{eq:V3_three_source}
\mathcal V_{\text{3SPDC}}(T,N_B)
\;=\;
\frac{2\sqrt{(1+V_A)(1+V_C)\,V_A\,V_B\,T}}
{K}\,,
\end{equation}
where $K = (1+V_C)V_A + V_B + V_C + T V_A V_B + (1-T)N_B V_B$.

\begin{figure}[t]
  \centering
  \includegraphics[width=1\linewidth]{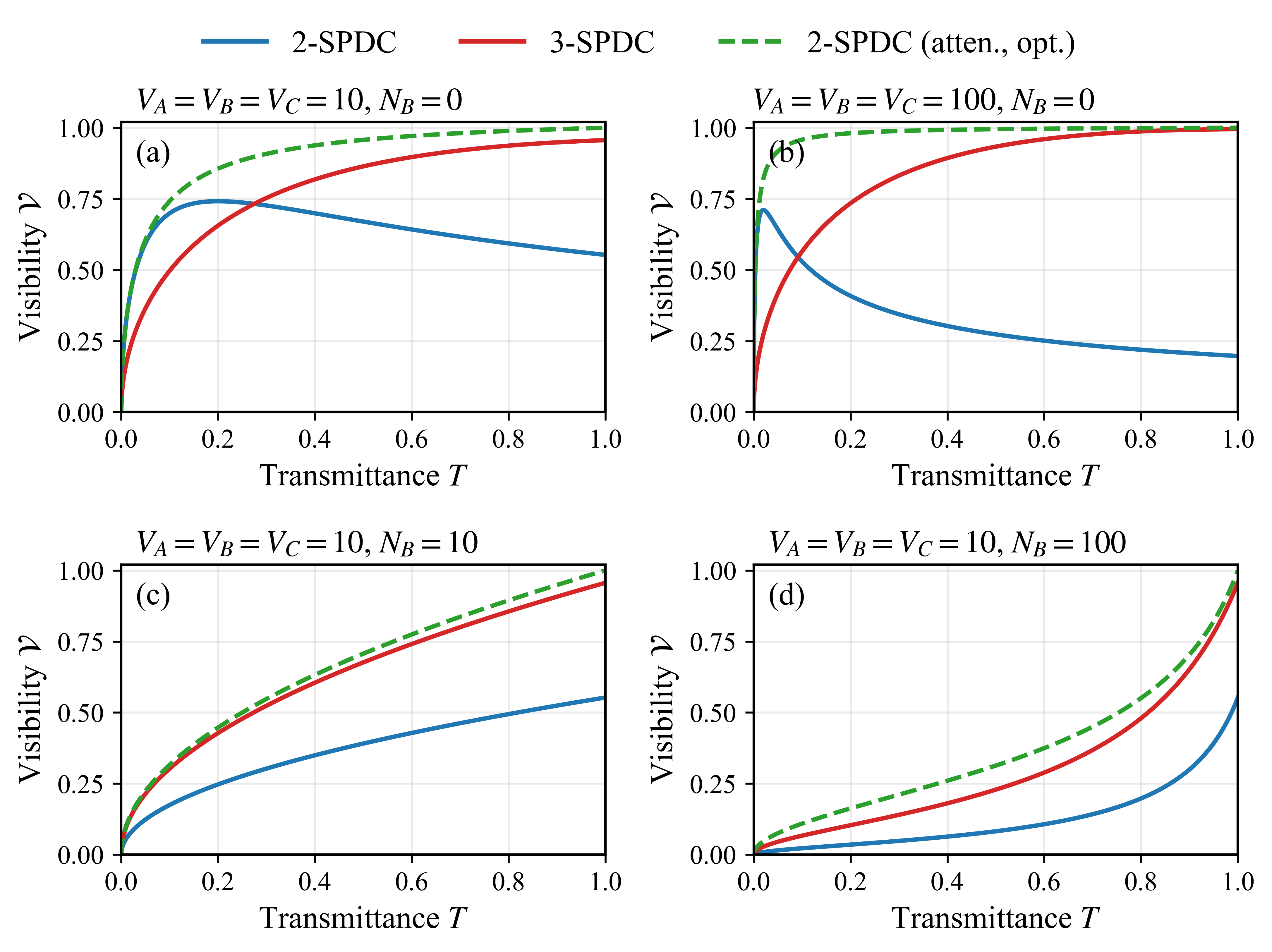}\hfill
  \caption{(Color online) Singles visibility $\mathcal{V}$ versus idler transmittance $T$ for three configurations: two crystals (2-SPDC, solid blue), three crystals with an added source in the $A$-signal arm (3-SPDC, solid red), and 2-SPDC with an \emph{optimized} attenuator in the $B$-signal arm (green dashed),
  computed from the analytic expressions in Sec.~II and App.~\ref{apa}. Panels show the effect of thermal background $N_B$ and High gain:
  (a) $V_A=V_B=V_C=10$; $N_B=0$,(b) $V_A=V_B=V_C=100$, $N_B=0$; (c) $V_A=V_B=V_C=10$, $N_B=10$; 
  (d) $V_A=V_B=V_C=10$, $N_B=100$.
  Thermal injection suppresses the 2-SPDC visibility except near $T\!\to\!1$ where the pedestal $(1-T)N_BV_B$ vanishes. Rebalancing the signal powers—either by adding a third SPDC (3-SPDC) or by optimally attenuating the stronger arm—drives the singles
  visibility toward the first-order coherence bound 
  $|g^{(1)}_{12}|=\sqrt{T(1+V_A)\big/(1+T V_A+(1-T)N_B)}$, restoring high contrast at large $T$ and in bright backgrounds.}
  \label{fig:3scheme_visibility}
\end{figure}

Figure~\ref{fig:3scheme_visibility} compares the singles visibility $\mathcal V$ predicted for three configurations—two-SPDC, optimally attenuated two-SPDC, and three-SPDC—under different gains and thermal backgrounds. The analytic trends of Eqs.~\eqref{eq:V2_atten_opt}–\eqref{eq:V3_three_source} show that both the attenuated and three-SPDC schemes recover the coherence-bound visibility even in bright thermal environments. As the thermal photon number $N_B$ increases, the unbalanced 2-SPDC visibility collapses except near $T\!\to\!1$, whereas rebalancing the signal powers (either by attenuation or by adding a third SPDC) maintains high-contrast fringes across a wide range of $T$.

In the low-gain regime $V_{A,B}\!\ll\!1$ under strong thermal injection $N_B\!\gg\!1$, the pedestal term $(1-T)N_BV_B$ in the denominator of Eq.~\eqref{eq:V_general_short} overwhelms the $O(\sqrt{V_A V_B})$ coherent term in the numerator, and the singles visibility collapses, $\mathcal V\!\to\!0$ for any fixed $T<1$ (and remains vanishingly small even as $T\!\to\!1$). In this regime we therefore switch to \emph{idler $(\hat{a}_{3}^{3})$-heralded} detection: conditioning on an idler $(\hat{a}_{3}^{3})$ click projects measurements onto the two-photon (pair) sector and removes the thermal pedestal, restoring high-contrast interference that is essentially insensitive to $N_B$ in the single-mode, zero-accidental limit. We develop this heralded scheme and its performance in the next section.

\section{Heralded Induced-Coherence Interferometry}

In the heralded measurement~\cite{sperling,Somaraju,MandelWolf1995}, an idler $(\hat{a}_{3}^{3})$ detection event acts as a projective filter that selects only those signal $(\hat{a}_{1}^{1},\hat{a}_{2}^{1})$ photons that are genuinely pair-correlated with an idler $(\hat{a}_{3}^{1})$ generated in crystal~A. Because only crystal~A’s idler $(\hat{a}_{3}^{1})$ traverses the object port, an idler $(\hat{a}_{3}^{3})$ click identifies an event in which the corresponding signal $(\hat{a}_{1}^{1})$ photon carries phase information about the object. Uncorrelated photons—those originating from the thermal background—have no joint correlation with the heralding event and therefore make no contribution to the conditional average. Operationally, the heralding process does not create new coherence; rather, it reveals the latent induced coherence already present in the entangled pair subspace by excluding all uncorrelated noise realizations. In this sense, heralding functions as a ``thermal projector'': it removes the additive thermal pedestal that degrades ordinary singles visibility and restores interference contrast determined solely by the coherent pair amplitude.

In singles detection ($\hat{a}_{2}^{2}$), the signal detector $F$ accumulates contributions from A-events (object-dependent interference), B-events (object-independent reference), and thermal photons (uncorrelated noise). As $N_B$ increases, the uncorrelated contributions wash out the interference visibility. By contrast, heralded detection conditions the signal $F$ counts on an idler ($\hat{a}_{3}^{3}$) click, giving the conditional expectation value
\begin{equation}
\langle n_S \rangle_{\text{cond}} = 
\frac{\langle n_I n_S \rangle}{\langle n_I \rangle},
\end{equation}
where only photons that are pair-correlated with the detected idler contribute. The phase-sensitive correlation between idler and signal fields includes coherent amplitudes from both SPDC sources,
\begin{equation}
\langle \hat{a}_I \hat{a}_S \rangle
= \langle \hat{a}_I \hat{S}_A \rangle + \langle \hat{a}_I \hat{S}_B \rangle \neq 0,
\end{equation}
while the thermal field is uncorrelated with the heralding idler and satisfies
\begin{equation}
\langle \hat{a}_I \hat{a}_S \rangle_{\mathrm{(thermal)}} = 0.
\end{equation}

Here $\hat{S}_A$ and $\hat{S}_B$ denote the signal output modes from crystals~A and~B, respectively, after their respective beam-splitter transformations defined in App.~\ref{apa}. Each SPDC source generates a signal--idler pair $(\hat{S}_j,\hat{a}_{3}^{j})$ with $j=A,B$, and the total signal field reaching detector~$F$ is the coherent superposition $\hat{a}_S=(\hat{S}_A+e^{i2\phi}\hat{S}_B)/\sqrt{2}$. The detected idler $\hat{a}_I\!\equiv\!\hat{a}_{3}^{3}$ contains indistinguishable contributions from both crystals, so its phase-sensitive correlation with the signal field separates naturally into two coherent amplitudes, $\langle \hat{a}_I \hat{S}_A\rangle$ and $\langle \hat{a}_I \hat{S}_B\rangle$, corresponding to the biphoton pathways through crystals~A and~B. In contrast, the thermal mode mixed through the object port is statistically independent of the SPDC fields and therefore yields $\langle \hat{a}_I \hat{a}_S\rangle_{\mathrm{(thermal)}}=0$.

Mathematically,
\begin{equation}
\begin{split}
|\langle \hat{a}_I \hat{a}_S \rangle|^2 &\neq 0 
\quad \text{for correlated (A,B) events},\\
|\langle \hat{a}_I \hat{a}_S \rangle|^2 &= 0 
\quad \text{for uncorrelated thermal events}.
\end{split}
\end{equation}
Consequently, the conditional statistics retain only the coherent superposition of A and B contributions, while the uncorrelated thermal background is projected out. 

\paragraph*{Physical meaning.}
Only crystal~A’s idler $\hat{a}_{3}^{2}$ passes through the object port before overlapping with idler $\hat{a}_{3}^{3}$ from crystal~B. Crystal~B’s idler bypasses the object and provides a reference path. When the two idler modes are aligned, they become indistinguishable, and heralding projects the measurement onto the two-photon subspace that carries no which-path information. In this regime, interference arises from the coherent superposition of the A (object-sensing) and B (reference) amplitudes, while all uncorrelated thermal photons are rejected.

The heralded singles intensity follows from the two-photon correlations in App.~\ref{apa} as
\begin{equation}
N^{(\mathrm{her})}_\pm(\phi)
  = \tfrac{1}{2}\Big[V_A + T V_A V_B 
  \pm 2\sqrt{T(1+V_A)V_A V_B}\cos(2\phi)\Big],
\end{equation}
giving the heralded visibility
\begin{equation}
\label{eq:Vherald-final}
\mathcal{V}_{\mathrm{herald}}
  = \frac{2\sqrt{T(1+V_A)V_A V_B}}
         {V_A +V_B+ T V_A V_B},
\end{equation}
which is independent of the thermal photon number~$N_B$.

Equation~\eqref{eq:Vherald-final} shows that the heralded visibility is insensitive to thermal noise: the additive pedestal $(1-T)N_BV_B$ that limited the singles contrast in Eq.~\eqref{eq:V_general_short} is absent. Conditioning on an idler ($\hat{a}_{3}^{3}$) click projects the measurement onto the two-photon subspace, retaining only events genuinely correlated through down-conversion in the two SPDC sources. For small gain ($V_A,V_B\!\ll\!1$), $\mathcal{V}_{\mathrm{herald}}$ increases monotonically with~$T$ and approaches the ideal ZWM limit at $T\!\to\!1$; at higher gain, stimulated emission in crystal~B causes a small imbalance that slightly reduces the contrast.

\begin{figure}[t]
\centering
\includegraphics[width=1\linewidth]{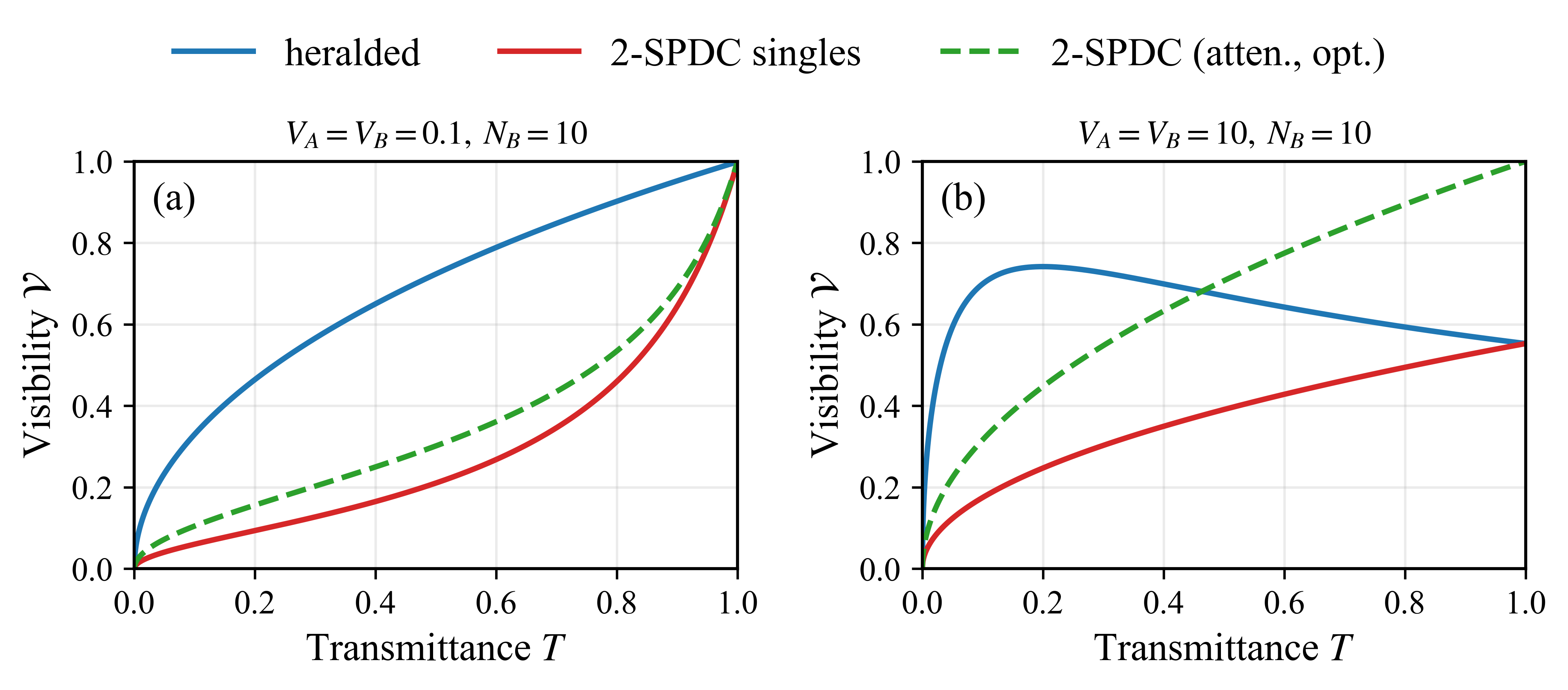}\hfill
\caption{(Color online) Visibility $\mathcal{V}$ versus idler transmittance $T$ for three detection schemes: heralded (solid blue, Eq.~\eqref{eq:Vherald-final}), two-SPDC singles (solid red, Eq.~\eqref{eq:V_general_short}), and two-SPDC with optimal attenuation (green dashed, $|g^{(1)}_{12}|$). Low gain: (a)~$V_A=V_B=0.1$, $N_B=10$: the thermal pedestal $(1-T)N_BV_B$ suppresses singles visibility over most of $T$, whereas heralding remains high and monotonic with~$T$. High gain: (b)~$V_A=V_B=10$, $N_B=10$: heralded visibility peaks at small $T$ and decreases slightly as stimulated imbalance ($\propto T V_A V_B$) grows.}
\label{fig:herald_vs_singles_nb10}
\end{figure}

\section{Unconditional and Conditional SNR}
\paragraph*{Why SNR matters in ZWM.}
The \emph{signal-to-noise ratio} (SNR) of the detected
photon-number difference between the two outputs of $S_1$
quantifies how reliably interference fringes can be resolved within a
finite acquisition time. For the two-crystal ZWM geometry with thermal
injection, maximizing over phase gives the \emph{unconditional} singles
SNR, $\mathrm{SNR}_{\text{2SPDC}}^{\max}(T)$, in
Eq.~(\ref{eq:SNR_2SPDC_max}). These SNR expressions directly determine
the integration time required to reach a given confidence level and show
how transmittance~$T$, parametric gains $(V_A,V_B,V_C)$, and thermal
brightness~$N_B$ trade off in practice.

\paragraph*{Unconditional SNR.}
In singles detection, both signal outputs contain additive thermal and
background contributions, so the photon-number–difference operator
$\hat N_-=\hat N_1-\hat N_2$ has an expected mean
$\langle\hat N_-\rangle\!\propto\!\mathcal V(T)\cos(2\phi)$ and variance
$\langle(\Delta\hat N_-)^2\rangle=\langle\hat N_1\rangle+\langle\hat
N_2\rangle$ (see App.~\ref{apb}). Maximizing over phase gives the peak
SNR,
\begin{equation}
\label{eq:SNR_2SPDC_max}
\mathrm{SNR}_{\text{2SPDC}}^{\max}(T,N_B)
=\frac{N_+^{\max}-N_+^{\min}}
{\sqrt{\langle(\Delta \hat N_-)^2\rangle}}
\end{equation}

\begin{equation}
\label{eq:SNR_2SPDC_max}
\mathrm{SNR}_{\text{2SPDC}}^{\max}(T,N_B)
=\frac{2\sqrt{T(1+V_A)V_A V_B}}
{\sqrt{V_A+V_B+T V_A V_B+(1-T)N_B V_B}}.
\end{equation}
For small $T$, this SNR scales linearly with~$T$, while at large $T$ it
saturates as $\langle\hat N_2\rangle\!\gg\!\langle\hat N_1\rangle$.
Importantly, the additive thermal term $(1-T)N_BV_B$ reduces the SNR even
when visibility remains finite, since thermal noise raises the shot-noise
floor.

\paragraph*{Conditional (heralded) SNR.}
Conditioning on an idler detection removes uncorrelated background
counts, confining the statistics to the two-photon subspace. The
heralded photon-number–difference SNR then follows from the correlated
pair amplitudes alone:
\begin{equation}
\label{eq:SNR-herald-LB}
\mathrm{SNR}_{\text{herald}}^{\max}(T)
=\frac{2\sqrt{T(1+V_A)V_A V_B}}
{\sqrt{V_A+T V_A V_B}}.
\end{equation}
This expression is \emph{independent} of the thermal photon number
$N_B$, confirming that heralding projects away the uncorrelated thermal
background. In the low-gain limit ($V_A,V_B\!\ll\!1$),
$\mathrm{SNR}_{\text{herald}}^{\max}\!\propto\!T$, matching the expected
single-pair scaling, while at high gain, stimulated emission in crystal
B again limits the attainable contrast at large~$T$.

\paragraph*{Physical interpretation.}
Equation~\eqref{eq:SNR_2SPDC_max} shows that in the presence of thermal
noise, the unconditional SNR degrades as $(1-T)N_B$ increases, reflecting
both signal attenuation and background-induced variance. By contrast,
Eq.~\eqref{eq:SNR-herald-LB} shows that the conditional SNR is
insensitive to $N_B$ and remains bounded only by the parametric gains.
Thus, heralding effectively reestablishes the quantum
shot-noise--limited scaling that would otherwise be buried under
thermal fluctuations.

\begin{figure}[t]
  \centering
  \includegraphics[width=1\linewidth]{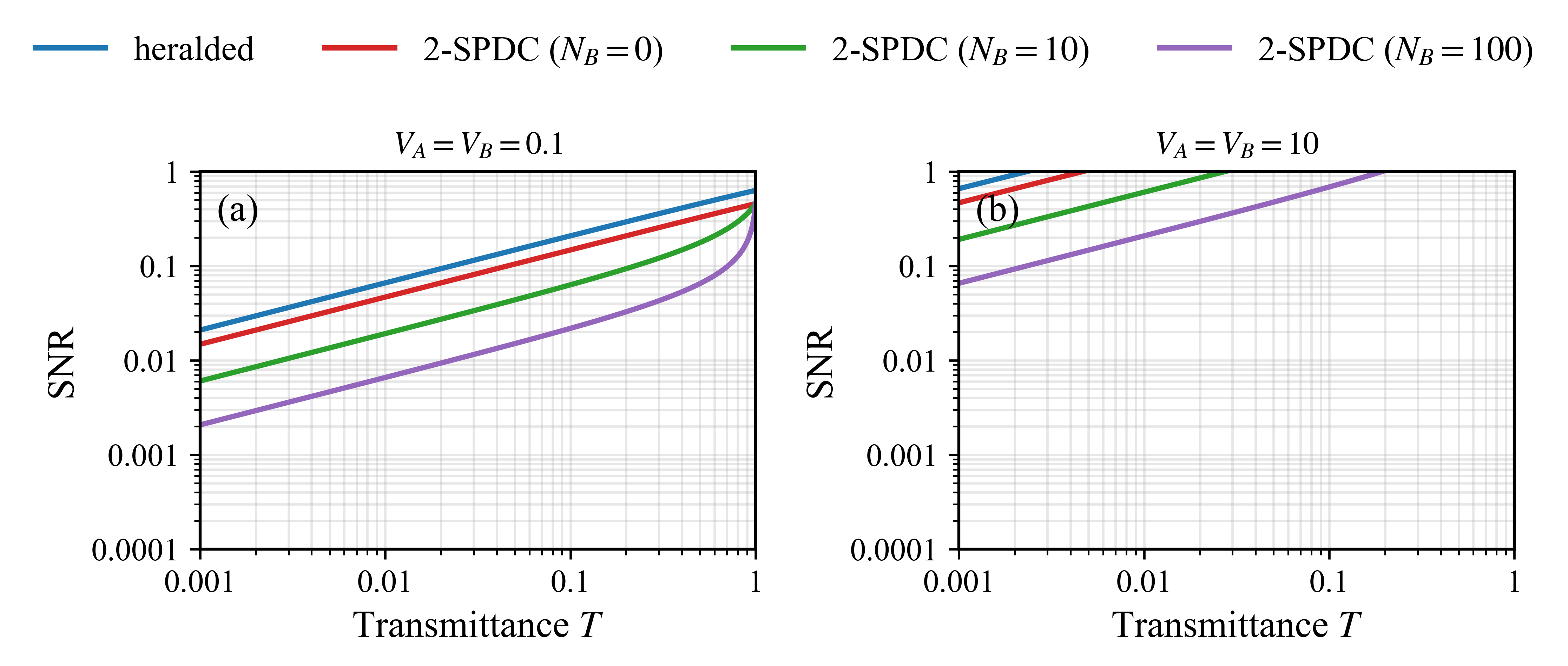}\hfill
  
  \caption{Log--log scaling of the photon–number–difference SNR versus
  idler transmittance~$T$ in the low gain (a) and high gain (b). Blue: \emph{heralded} SNR, independent of
  thermal background $N_B$ [Eq.~\eqref{eq:SNR-herald-LB}]. Colored
  curves: \emph{unconditional} two-SPDC SNR for
  $N_B\!\in\!\{0,1,10,100\}$ [Eq.~\eqref{eq:SNR_2SPDC_max}]. For small
  $T$, all curves scale linearly with~$T$; as $N_B$ increases, the
  unconditional SNR decreases while the heralded SNR remains constant.
  At high $T$, both converge as the thermal pedestal
  $(1-T)N_BV_B\!\to\!0$.}
  \label{fig:snr-scaling-loglog}
\end{figure}

Figure~\ref{fig:snr-scaling-loglog} compares the log--log scaling of
unconditional and heralded SNR versus idler transmittance~$T$ across
different gains and thermal backgrounds. For small~$T$, all SNRs scale
linearly with~$T$, but as $N_B$ increases, the unconditional (colored)
curves drop sharply while the heralded (blue) curve remains unchanged.
At high transmittance ($T\!\to\!1$) the curves converge, since the
thermal pedestal vanishes. Across all $T$ and gain regimes, the heralded
SNR stays at or above the unconditional SNR whenever $N_B>0$, clearly
demonstrating its robustness against thermal noise.

\section{Conclusion}

We presented a quantum-optical analysis of induced-coherence interferometry in the Zou–Wang–Mandel (ZWM) geometry with a thermally seeded idler port. Using Bogoliubov transformations, we derived closed-form expressions for singles intensities, first-order coherence, visibility, and signal-to-noise ratio (SNR) across low- and high-gain regimes, identifying how the additive pedestal $\propto(1-T)N_BV_B$ suppresses singles contrast.

When thermal backgrounds are present, visibility can be recovered \emph{without coincidences} through two passive rebalancing strategies: (i) \emph{optimal attenuation} of the stronger signal arm to equalize intensities, and (ii) a \emph{three-SPDC (3-SPDC)} configuration that restores balance at higher gain. Both approaches recover singles visibility up to the coherence limit $|g^{(1)}_{12}|$.
In the strongly thermal, low-gain regime where singles still collapse, \emph{idler-heralded} detection removes uncorrelated thermal photons, yielding visibility and SNR that are independent of $N_B$. Heralding thus provides coincidence-level background rejection using only singles detection, making it attractive for quantum imaging or sensing under bright ambient conditions.

Overall, these results establish attenuation-balanced and heralded induced-coherence interferometry as practical, noise-resilient sensing schemes that extend induced-coherence techniques into thermally bright spectral bands (mid-IR, THz, microwave).

The robustness of the heralded induced-coherence scheme can be intuitively understood by analogy to ghost imaging. In ghost imaging with SPDC sources~\cite{Pittman1995,Shapiro2012,Valencia2005}, an image that is invisible in either singles stream emerges only in the conditional correlations between a spatially resolving signal detector and a bucket idler detector. Thermal photons or stray light entering the idler arm do not correlate with the signal and are therefore rejected by the coincidence measurement. Heralded induced-coherence interferometry performs an analogous role in the temporal domain: an idler click projects the signal statistics onto the correlated pair subspace, eliminating the thermal pedestal that collapses the singles visibility. In this sense, heralding acts as a ``thermal projector'' for induced coherence, directly paralleling how correlation-based ghost imaging rejects uncorrelated background to recover image contrast.


\begin{acknowledgments}
L.~T. thanks Kanad Sengupta for helpful discussions. 
L.~T. and C.~M.~C. acknowledge support from the Office of the Principal Scientific Adviser to the Government of India under Project No.~Prn.SA/QSim/2020. B. V acknowledges the support from the Institute of Mathematical Sciences, Chennai. 
\end{acknowledgments}

\appendix
\section{Single-Mode Analysis of the ZWM Interferometer and the Calculation of Visibility}\label{apa}
	We analyze the setup shown in Fig.~\ref{2psdc} in the Heisenberg picture \cite{Kolobov2017}. 
Each mode $i=1,2,3,4$ of the nonlinear interferometer is described by annihilation and creation operators 
$\hat{a}_{i}$ and $\hat{a}_{i}^{\dagger}$ satisfying the standard bosonic commutation relation 
$[ \hat{a}_{i},\hat{a}_{j}^{\dagger}]=\delta_{ij}$. 
The mean photon number in mode~$i$ is given by the expectation value 
$\langle \hat{N}_{i} \rangle = \langle \hat{a}_{i}^{\dagger} \hat{a}_{i} \rangle$. 
This operator-based description is standard in quantum optics for representing the state of light in each mode.

    The annihilation operators of the input modes to crystal A are $\hat{a}_{1}$ and $\hat{a}_{3}$, and the corresponding annihilation operators of the output modes from crystal A are $\hat{a}_{1}^{1}$ and $\hat{a}_{3}^{1}$ (the subscript indicates the input mode, and the superscript indicates the output mode as shown in Fig.\ref{2psdc}). We assume that the pump is a classical field that remains undepleted throughout the process. The output modes are related to the input modes by the Bogoliubov transformation which is written as
	\begin{equation}\label{bo1}
		\begin{split}
	\hat{a}_{1}^{1}&=u_{A}\hat{a}_{1}+v_{A}\hat{a}_{3}^{\dagger}\\
		\hat{a}_{3}^{1}&=u_{A}\hat{a}_{3}+v_{A}\hat{a}_{1}^{\dagger},
		\end{split}
	\end{equation}
	such that $\vert u_{A}\vert^2-\vert v_{A}\vert^2=1$. In general,  $u_A$ and $v_A$ are complex numbers; $|u_{A}|$ and $|v_{A}|$ can be represented by hyperbolic functions: $|u_{A}| = \cosh{(\chi_{A})}$ and $|v_{A}|= \sinh{(\chi_{A})}$, where $\chi_{A}$ is the coupling parameter to the nonlinear medium. We define $U_{A} \equiv |u_{A}|^{2}$ and $V_{A} \equiv |v_{A}|^{2}$ such that $U_{A} - V_{A} = \cosh^{2}{(\chi_A)}-\sinh^{2}{(\chi_A)}=1 $.

\par
    The transmittance in the idler $\hat{a}_{3}^{1}$ modes between crystals $A$ and $B$, combined with the thermal background noise $\hat{a}_{4}$ [see Fig.\ref{2psdc}] with mean thermal photon number $N_B=\langle \hat{a}^{\dagger}_{4}\hat{a}_{4} \rangle$, results in the transformation.
	\begin{equation} \label{A2}
		\begin{split}
			\hat{a}_{3}^{2}&=t_{o}\hat{a}_{3}^{1}+r_{o}\hat{a}_{4},\\
			\hat{a}_{4}^{1}&=t_{o}\hat{a}_{4}-r_{o}\hat{a}_{3}^{1},
		\end{split}
	\end{equation}
	where $\hat{a}_{3}^{2}$ and $\hat{a}_{4}^{1}$ are the annihilation operators of the output modes from the object $S_1$. We assume that $t_o ,r_o\in \rm{R}$ that satisfy the relation: $\vert t_{o}\vert^2+\vert r_{o}\vert^2=1:=T+R$, where $|t_{o}\vert^2 \equiv T$ and $\vert r_{o}\vert^2 \equiv R$.

On substituting for $\hat{a}_{3}^{1}$ from Eq. \ref{bo1} in Eq. \ref{A2}, we obtain
	\begin{equation}\label{a32}
		\hat{a}_{3}^{2}=t_{o}v_{A}\hat{a}_{1}^{\dagger}+t_{o}u_{A}\hat{a}_{3}
		+r_{o}\hat{a}_{4}.
	\end{equation}

	The two idler modes $\hat{a}_{3}^{1}$ and $\hat{a}_{3}^{3}$ are mode matched through alignment which is realized by seeding crystal B with the idler photon from crystal $A$ (represented by mode $\hat{a}_{3}^{1}$). The intensity transmittance $T$ of the object $S$ and the thermal background noise determine the effectiveness of alignment of the idler beam from crystal $A$ to the idler beam from crystal $B$ (represented by mode $\hat{a}_{3}^{3}$). The output modes from the second crystal $B$ are related to the input modes $\hat{a}_{2}$ and $\hat{a}_{3}^{2}$ by the Bogoliubov transformation
	\begin{equation}
		\begin{split}
			\hat{a}_{2}^{1}&=u_{B}\hat{a}_{2}+v_{B}\hat{a}_{3}^{2\dagger},\\
			\hat{a}_{3}^{3}&=u_{B}\hat{a}_{3}^{2}+v_{B}\hat{a}_{2}^{\dagger},
		\end{split}
	\end{equation}
	such that $\vert u_{B}\vert^2-\vert v_{B}\vert^2=1$, $|u_{B}|=\cosh{(\chi_{B})}$ and $|v_{B}|=\sinh{(\chi_{B})}$.
	
	The signal mode $\hat{a}_{2}^{1}$ can be expressed in terms of the input modes using Eq.\ref{a32} which then gives us
	\begin{equation}\label{a21}
 \hat{a}_{2}^{1}=t_{o}v_{A}^{*}v_{B}\hat{a}_{1}+u_{B}\hat{a}_{2}+t_{o}u_{A}^{*}v_{B}\hat{a}_{3}^{\dagger}+r_{o}v_{B}\hat{a}_{4}^{\dagger},
	\end{equation}
	where $*$ describes the complex conjugate. 
	
	The two signal photons from both crystals $A$ and $B$ in modes $\hat{a}_{1}^{1}$ and $\hat{a}_{2}^{1}$, respectively, pass through the  $50:50$ beam splitter $S_1$ whose output modes are given by
    
     \begin{equation}
		\begin{split}
			\hat{a}_{2}^{2}&=(\hat{a}_{2}^{1}+\hat{a}_{1}^{1})/\sqrt{2}\\
			\hat{a}_{1}^{2}&=(\hat{a}_{2}^{1}-\hat{a}_{1}^{1})/\sqrt{2}.
		\end{split}
	\end{equation}

	The two output signal modes $\hat{a}_{2}^{1}$ and $\hat{a}_{1}^{1}$ from the beam splitter can be expressed in terms of the input modes using Eq.\ref{a21} and Eq.\ref{bo1}, which yields
	\begin{equation}
		\begin{split}
			\hat{a}_{2,1}^{2,2}=&( t_{o}^{}v_{A}^{*}v_{B}^{} \pm u_{A})\hat{a}_{1}+( t_{o}^{}u_{A}^{*}v_{B}\pm v_{A})\hat{a}_{3}^{\dagger}\\
			&+ u_{B}\hat{a}_{2}+ r_{o}^{}v_{B}\hat{a}_{4}^{\dagger}.
		\end{split}
	\end{equation}
	
	Following this, we can calculate the mean photon numbers 
	$\hat{N}_{2,1}^{2,2}$ in the two output ports of the beam splitter which are given by
	\begin{equation}\label{nb}
		\begin{split}
			\hat{N}_{2,1}^{2,2}=&\langle\ \hat{a}_{2,1}^{2,2,\dagger}\hat{a}_{2,1}^{2,2}\ \rangle\\
			= &\frac{1}{2}\big[V_{A}+TV_{A}V_{B}
			 +V_{B}+(1-T)N_BV_{B} \\
			&\pm 2\sqrt{T(1+V_{A})V_{A}V_{B}}\cos{(2\phi)}\big].
		\end{split}
	\end{equation}

	Since the crystals are not seeded, we assume a vacuum input where $\langle\hat{a}^{\dagger}_{i}\hat{a}_{i}\rangle=0$ for all $i=1,2,3,4,5$. Given this assumption and the definition,
	\begin{equation}
		t_{o}u_{A}v_{A} v_{B}^{*}=\sqrt{TU_{A}V_{A}V_{B}}\exp{(i2\phi)},
	\end{equation}
	where  $u_{i}$ and $v_{i}$ are the complex parameters of the Bogoliubov transformation, satisfying $U_{i}-V_{i}=|u_i|^2-|v_i|^2=1$.

\section{Variance of the photon-number difference}\label{apb}

We define the photon-number difference operator as
\begin{equation}
\hat{N}_- = \hat{N}_1 - \hat{N}_2 = 
\hat{a}_1^{2 \, \dagger} \hat{a}_1^{2} - 
\hat{a}_2^{2 \, \dagger} \hat{a}_2^{2}.
\end{equation}

The corresponding variance is
\begin{equation}
\langle (\Delta \hat{N}_-)^2 \rangle 
= \langle \hat{N}_-^2 \rangle - \langle \hat{N}_- \rangle^2.
\end{equation}

Expanding the square gives
\begin{equation}
\langle \hat{N}_-^2 \rangle 
= \langle \hat{N}_1^2 \rangle 
+ \langle \hat{N}_2^2 \rangle 
- 2\langle \hat{N}_1 \hat{N}_2 \rangle.
\end{equation}

To evaluate $\langle \hat{N}_i^2 \rangle$ for each mode, we recall that for a bosonic mode 
$\hat{a}$ with number operator $\hat{N}=\hat{a}^\dagger \hat{a}$,
\begin{equation}
\hat{N}^2 
= \hat{a}^\dagger \hat{a}^\dagger \hat{a} \hat{a} 
+ \hat{a}^\dagger \hat{a}
= \hat{N}(\hat{N}-1) + \hat{N}.
\end{equation}
Taking the expectation value yields
\begin{equation}
\langle \hat{N}^2 \rangle 
= \langle \hat{N}(\hat{N}-1)\rangle + \langle \hat{N}\rangle.
\end{equation}
For thermal or Poissonian (shot-noise-limited) light fields where intensity fluctuations are uncorrelated between different modes, the photon statistics satisfy 
$\langle \hat{N}(\hat{N}-1)\rangle = \langle \hat{N}\rangle^2$, so that
\begin{equation}
\boxed{\langle \hat{N}^2 \rangle = \langle \hat{N}\rangle^2 + \langle \hat{N}\rangle.}
\end{equation}

Similarly, for statistically independent modes $1$ and $2$, the cross term factorizes as
\begin{equation}
\boxed{\langle \hat{N}_1 \hat{N}_2\rangle 
= \langle \hat{N}_1\rangle \langle \hat{N}_2\rangle.}
\end{equation}

Substituting these relations into Eq.~(B3) gives
\begin{equation}
\begin{aligned}
\langle (\Delta \hat{N}_-)^2 \rangle
&= (\langle \hat{N}_1\rangle^2 + \langle \hat{N}_1\rangle)
 + (\langle \hat{N}_2\rangle^2 + \langle \hat{N}_2\rangle)
 - 2\langle \hat{N}_1\rangle \langle \hat{N}_2\rangle \\
&\quad - (\langle \hat{N}_1\rangle - \langle \hat{N}_2\rangle)^2 \\
&= \langle \hat{N}_1\rangle + \langle \hat{N}_2\rangle.
\end{aligned}
\end{equation}

This result shows that the variance of the photon-number difference equals the sum of the mean photon numbers at the two outputs, corresponding to the shot-noise limit for uncorrelated detection statistics.

\section{Heralded (conditional) visibility: mode-matched condition}\label{apc}

We work in the Heisenberg picture with vacuum inputs $\hat a_{1,2,3}$ and an idler object with amplitude transmissivity $t_o$ and $T=|t_o|^2$ between crystals $A$ and $B$. The detected idler mode is $\hat b_I\equiv \hat a_3^{\,3}$ and the detected signal mode at the ``$+$'' port is $\hat b_S\equiv \hat c_+=(\hat a_2^{\,1}+\hat a_1^{\,1})/\sqrt{2}$. Crystals $X\in\{A,B\}$ are two-mode squeezers with $u_X=\cosh r_X$, $v_X=\sinh r_X\,e^{i\theta_X}$ and $U_X:=|u_X|^2=1+V_X$, $V_X:=|v_X|^2$. We assume \emph{mode-matched heralding}: in the conditional (heralded) analysis, the detected idler mode is mode-matched to the SPDC idler and statistically independent of the thermal background. This ensures that thermal photons—while present in the field—contribute only as additive intensity noise and do not affect the phase-sensitive correlation $\langle \hat{a}_I \hat{a}_S\rangle$ entering the conditional expectation value $\langle n_I n_S\rangle/\langle n_I\rangle$.

\paragraph*{Heisenberg maps (mode matched).}
\begin{align}
\hat a_1^{\,1}&=u_A \hat a_1+v_A \hat a_3^\dagger,\qquad
\hat a_3^{\,2}=t_o u_A \hat a_3+t_o v_A \hat a_1^\dagger,\\
\hat a_2^{\,1}&=u_B \hat a_2+v_B(\hat a_3^{\,2})^\dagger
=u_B \hat a_2+t_o v_A^* v_B\,\hat a_1+t_o u_A^* v_B\,\hat a_3^\dagger,\\
\hat c_+&=\frac{1}{\sqrt{2}}\!\left[(t_o v_A^* v_B+u_A)\hat a_1+(t_o u_A^* v_B+v_A)\hat a_3^\dagger+u_B \hat a_2\right],\\
\hat a_3^{\,3}&=u_B \hat a_3^{\,2}+v_B \hat a_2^\dagger
=u_B(t_o u_A \hat a_3+t_o v_A \hat a_1^\dagger)+v_B \hat a_2^\dagger.
\end{align}
Define the interference phase by
\begin{equation}
t_o u_A v_A v_B^*=\sqrt{T U_A V_A V_B}\,e^{i2\phi}.
\end{equation}

	\paragraph*{POVM conditioning.}
	Let the idler be measured by an on/off detector with POVM elements\cite{sperling,Somaraju,MandelWolf1995}
	\[
	E_{\rm click}=\mathbb I-:\!\exp(-\eta_I \hat n_I-\nu)\!:,\quad
	E_{\rm noclick}=:\!\exp(-\eta_I \hat n_I-\nu)\!:,
	\]
	where $\eta_I$ is the quantum efficiency, $\nu$ is the mean dark counts, and
	$\hat n_I=\hat a_3^{\,3\dagger}\hat a_3^{\,3}$ is the idler photon number in the detected mode.
	For any signal observable $O_S$ (we take $O_S=\hat n_S=\hat c_+^\dagger\hat c_+$),
	the idler-\emph{conditioned} average is
	\begin{equation}
  \langle O_S\rangle_{\mathrm{cond}}
  = \frac{\operatorname{Tr}\!\left[\rho\, (E_{\mathrm{click}}\!\otimes\!O_S)\right]}
         {\operatorname{Tr}\!\left[\rho\, (E_{\mathrm{click}}\!\otimes\!\mathbb{I})\right]} \,.
  \label{eq:general-POVM-cond}
\end{equation}
	In the low-brightness regime ($\eta_I\!\langle \hat n_I\rangle\ll1$, $\nu\ll1$),
	expand $E_{\rm click}\approx \eta_I \hat n_I+\nu$ to get
	\begin{equation}
		\boxed{\;
			\langle \hat n_S\rangle_{\rm cond}
			\;\approx\;
			\frac{\eta_I\langle \hat n_I \hat n_S\rangle+\nu\langle \hat n_S\rangle}
			{\eta_I\langle \hat n_I\rangle+\nu}
			\;\xrightarrow[\nu\to 0]{}\;
			\frac{\langle \hat n_I \hat n_S\rangle}{\langle \hat n_I\rangle}\,.
			\;}
		\label{eq:coins-over-singles}
	\end{equation}
	
	\paragraph*{Fourth order factorization (Gaussian inputs).}
	All inputs are zero-mean Gaussian (vacuum in $a_{1,2,3}$, thermal in $a_4$),
	so Wick/Isserlis yields
	\begin{equation}
		\langle \hat n_I \hat n_S\rangle
		= \langle \hat n_I\rangle\langle \hat n_S\rangle
		+ \big|\langle \hat a_3^{\,3}\hat c_{+}\rangle\big|^2
		+ \big|\langle \hat a_3^{\,3}\hat c_{+}^\dagger\rangle\big|^2.
		\label{eq:wick-herald}
	\end{equation}

For any zero-mean Gaussian state (vacuum inputs), Wick/Isserlis implies
\begin{equation}
\label{eq:cond-master}
\begin{split}
\langle \hat n_S\rangle_{\rm cond}(\phi)
=\frac{\langle \hat n_I \hat n_S\rangle}{\langle \hat n_I\rangle}
=\langle \hat n_S\rangle
+\frac{\big|\langle \hat b_I \hat b_S\rangle\big|^2+\big|\langle \hat b_I \hat b_S^\dagger\rangle\big|^2}{\langle \hat n_I\rangle}.
\end{split}
\end{equation}
In our geometry $\langle \hat b_I \hat b_S^\dagger\rangle=0$.

\paragraph*{Singles and covariance (mode matched).}

\begin{equation}
\begin{split}
\langle \hat n_I\rangle
&=\langle \hat a_3^{\,3\dagger}\hat a_3^{\,3}\rangle
=U_B\,T\,V_A+V_B,\\
\langle \hat n_S\rangle
&=\frac{1}{2}\!\big[V_A+V_B+T V_A V_B\\
&+2\sqrt{T(1+V_A)V_A V_B}\cos(2\phi)\big],\\
\big|\langle \hat b_I \hat b_S\rangle\big|^2
&=\frac{U_B}{2}\;T U_A\!\big[T U_A V_B+V_A\\
&+2\sqrt{T U_A V_A V_B}\cos(2\phi)\big].
\label{eq:cov2-mm}   
\end{split}
\end{equation}

\paragraph*{Conditional fringe and visibility.}
Inserting \eqref{eq:cov2-mm} into \eqref{eq:cond-master} yields
\begin{equation}
\langle \hat n_S\rangle_{\rm cond}(\phi)
=\bar N_{\rm cond}+A_{\rm cond}\cos(2\phi),
\end{equation}
with the DC and AC parts
\begin{align}
\bar N_{\rm cond}
&=\frac{1}{2}\big[V_A+V_B+T V_A V_B\big]\\
&+\frac{U_B}{2\,[U_B T V_A+V_B]}\Big[V_B\,(T U_A)^2+T U_A V_A\Big],\\
A_{\rm cond}
&=\sqrt{T(1+V_A)V_A V_B}\\
&+\frac{U_B}{U_B T V_A+V_B}\;T U_A\sqrt{T U_A V_A V_B}.
\end{align}
The heralded (conditional) visibility is
\begin{equation}
\label{eq:Vherald-general}
\mathcal V_{\rm herald}=\frac{A_{\rm cond}}{\bar N_{\rm cond}}.
\end{equation}

\paragraph*{Pair-heralding (low-brightness) limit.}
Operationally, conditioning projects onto the two-photon sector. Writing the two interfering signal contributions with weights
\[
W_A=V_A,\quad W_B=V_B+T V_A V_B,\quad
|\Gamma|=\sqrt{T(1+V_A)V_A V_B},
\]
the universal two-beam formula $\mathcal V=2|\Gamma|/(W_A+W_B)$ gives
\begin{equation}
\label{eq:Vherald-final}
\mathcal V_{\rm herald}
\;\xrightarrow[\text{pair-heralding}]{}
\frac{2\sqrt{T(1+V_A)\,V_A\,V_B}}{V_A+V_B+T V_A V_B}\,.
\end{equation}
Compared with the unconditioned singles visibility (which contains the additive thermal pedestal $(1-T)N_B V_B$), Eq.~\eqref{eq:Vherald-final} is \emph{independent of $N_B$} under mode-matched heralding: the idler click removes the thermal background from the observed fringe.

\section{Signal-to-Noise Ratio in a Two-SPDC Quantum Interferometer}

We now specialize to the standard two-crystal ZWM geometry with thermal injection in the idler path. The singles at the two output ports of the final $50{:}50$ beam splitter(cf. App.~\ref{apa}).
Define the background and the coherent modulation
\begin{equation}
\begin{split}
\bar N=&\tfrac{1}{2}\!\left[V_A+V_B+T V_A V_B+(1-T)\,N_B V_B\right],\\
\Delta N=&\sqrt{T(1+V_A)V_A V_B}.
\end{split}
\end{equation}
Then \( \langle \hat N_+\rangle=\bar N+\Delta N\cos(2\phi)\) and
\( \langle \hat N_-\rangle=\bar N-\Delta N\cos(2\phi)\).
We take the difference operator \(\hat D=\hat N_+-\hat N_-\). Its mean is
\begin{equation}
\label{eq:mean_diff_2SPDC}
\langle \hat D\rangle \;=\; 2\,\Delta N\,\cos(2\phi)
\;=\;2\sqrt{T(1+V_A)V_A V_B}\,\cos(2\phi).
\end{equation}
Under the same Gaussian/shot-noise approximation used above,
\(\mathrm{Var}(\hat D)=\langle \hat N_+\rangle+\langle \hat N_-\rangle=2\bar N\).
Hence the signal-to-noise ratio is
\begin{equation}
\label{eq:SNR_2SPDC_general}
\begin{split}
\mathrm{SNR}_{\text{2SPDC}}(\phi)\;=\;&
\frac{\langle \hat D\rangle^2}{\mathrm{Var}(\hat D)}\\
\;=\;&
\frac{4\,T(1+V_A)\,V_A\,V_B\,\cos^2(2\phi)}
{V_A+V_B+T V_A V_B+(1-T)\,N_B V_B}\,.
\end{split}
\end{equation}
The SNR is maximized at constructive phase \((2\phi=0\!\!\mod 2\pi)\), yielding
\begin{equation}
\label{eq:SNR_2SPDC_max}
\mathrm{SNR}_{\text{2SPDC}}^{\max}(T)
\;=\;
\frac{4\,T(1+V_A)\,V_A\,V_B}
{V_A+V_B+T V_A V_B+(1-T)\,N_B V_B}\,.
\end{equation}
In the vacuum limit of the idler port \((N_B=0)\), Eq.~\eqref{eq:SNR_2SPDC_max} reduces to
\(\mathrm{SNR}^{\max}=4\,T(1+V_A)\,V_A V_B /(V_A+V_B+T V_A V_B)\).

\section{Heralded SNR in the two-SPDC interferometer (mode matched)}
\label{app:snr-herald-2spdc}

From Sec.~\ref{apc}, the conditional singles at the two
signal outputs (given an idler click) are
\begin{equation}
\label{eq:NpmCond}
\langle \hat N_\pm\rangle_{\rm cond}(\phi)
=\bar N_{\rm cond}\pm A_{\rm cond}\cos(2\phi),
\end{equation}
with the mode-matched DC and AC parts
\begin{align}
\bar N_{\rm cond}
&=\frac{1}{2}\big[V_A+V_B+T V_A V_B\big]\\
&+\frac{U_B}{2\,[U_B T V_A+V_B]}\Big[V_B\,(T U_A)^2+T U_A V_A\Big],
\label{eq:NbarCondMM-app}\\
A_{\rm cond}
&=\sqrt{T(1+V_A)V_A V_B}\\
&+\frac{U_B}{U_B T V_A+V_B}\;T U_A\sqrt{T U_A V_A V_B}.
\label{eq:ACondMM-app}
\end{align}
(Here $U_X=1+V_X$, $T=|t_o|^2$, and the phase is set by
$t_o u_A v_A v_B^*=\sqrt{T U_A V_A V_B}\,e^{i2\phi}$.)

Define the difference operator $\hat D_{\rm cond}=\hat N_+ - \hat N_-$.
Its conditional mean is
\begin{equation}
\langle \hat D_{\rm cond}\rangle
=2\,A_{\rm cond}\cos(2\phi).
\end{equation}
Under the same Gaussian/shot-noise approximation used for the
unconditioned case, the conditional variance is well-approximated by
\begin{equation}
\mathrm{Var}(\hat D_{\rm cond})
\simeq \langle \hat N_+\rangle_{\rm cond}
      + \langle \hat N_-\rangle_{\rm cond}
= 2\,\bar N_{\rm cond}.
\end{equation}
Hence the \emph{heralded} signal-to-noise ratio is
\begin{equation}
\label{eq:SNR-herald-general}
\mathrm{SNR}_{\rm herald}(\phi)
=\frac{\langle \hat D_{\rm cond}\rangle^2}{\mathrm{Var}(\hat D_{\rm cond})}
=\frac{2\,A_{\rm cond}^2\,\cos^2(2\phi)}{\bar N_{\rm cond}}\,.
\end{equation}

It is maximized at constructive phase $(2\phi=0\bmod 2\pi)$:
\begin{equation}
\label{eq:SNR-herald-max}
\mathrm{SNR}_{\rm herald}^{\max}
=\frac{2\,A_{\rm cond}^2}{\bar N_{\rm cond}}\,,
\quad
\text{with $A_{\rm cond}$ and $\bar N_{\rm cond}$.}
\end{equation}

\paragraph*{Pair-heralding (low-brightness) limit.}
In the operational two-photon limit, the conditional fringe takes the
universal two-beam form with
\[
W_A=V_A,\quad W_B=V_B+T V_A V_B,\quad
|\Gamma|=\sqrt{T(1+V_A)V_A V_B}.
\]
Thus
\(
\bar N_{\rm cond}\to \tfrac{1}{2}(W_A+W_B)
=\tfrac{1}{2}\big[V_A+V_B+T V_A V_B\big]
\)
and
\(
A_{\rm cond}\to |\Gamma|
=\sqrt{T(1+V_A)V_A V_B}.
\)
Inserting these into \eqref{eq:SNR-herald-general} gives the compact result
\begin{equation}
\label{eq:SNR-herald-LB}
\mathrm{SNR}_{\rm herald}^{\max}
\;\xrightarrow[\text{pair-heralding}]{}\;
\frac{4\,T(1+V_A)\,V_A\,V_B}{V_A+V_B+T V_A V_B}\,,
\end{equation}
which is \emph{independent of the thermal photon number $N_B$} under
mode-matched heralding (the idler click projects out the thermal pedestal).

\nocite{*}

\bibliography{apssamp}

\end{document}